\documentclass[12pt]{article}
\usepackage{amssymb,latexsym,epsfig,cite}

\newcommand{\be}{\begin{equation}}
\newcommand{\ee}{\end{equation}}
\def\bea{\begin{eqnarray}}
\def\eea{\end{eqnarray}}
\begin{document}

\thispagestyle{empty}

\begin{center}
{\Large \bf Gamow-Siegert functions and Darboux-deformed short
range potentials\footnote{NOTICE: this is the author's version of
a work that was accepted for publication in Annals of Physics.
Changes resulting from the publishing process, such as peer
review, editing, corrections, structural formatting, and other
quality control mechanisms may not be reflected in this document.
Changes may have been made to this work since it was submitted for
publication. A definitive version was subsequently published in
{\em Annals of Physics} {\bf 323} (2008) 1397-1414
[DOI:10.1016/j.aop.2007.11.002]}}
\end{center}


\begin{center}
Nicol\'as Fern\'andez-Garc\'{\i}a and Oscar Rosas-Ortiz\\[2ex]
{\footnotesize \it Departamento de F\'{\i}sica, Cinvestav, AP 14-740,
07000 M\'exico~D~F, Mexico}
\end{center}

\begin{abstract}
\noindent
Darboux-deformations of short range one-dimensional potentials are
constructed by means of Gamow-Siegert functions (resonance
states). Results include both Hermitian and non-Hermitian short
range potentials which are exactly solvable. As illustration, the
method is applied to square wells and barriers for which the
transmission coefficient is a superposition of Fock-Breit-Wigner
distributions. Resonance levels are calculated in the long
lifetime limit by means of analytical and numerical approaches.
The new complex potentials behave as an optical device which both
refracts and absorbs light waves.
\end{abstract}

\section{Introduction}
\label{int}

Gamow-Siegert functions are solutions of the Schr\"odinger
equation associated to complex eigenvalues and fulfilling purely
outgoing conditions \cite{Gam28,Sie39}. Although these functions
are not square-integrable, they are used to model physical
phenomena as resonances which, incidentally, are defined
experimentally better than theoretically. Some approaches extend
the formalism of quantum theory so that resonant states can be
defined in a precise form. For instance, in the formulation of the
rigged Hilbert space, Gamow vectors are studied in the framework
of the nuclear spectral theorem \cite{Boh89,Civ04}. Resonances are
also connected with the point spectrum of complex-scaled
Hamiltonians \cite{Agu71}.

In a different context, complex eigenvalues of Hermitian
Hamiltonians have been used to implement Darboux (supersymmetric)
transformations in quantum mechanics
\cite{Can98,Sam06a,Fer03,Ros03} (see also the discussion on
`atypical models' in \cite{Mie04}). The transformed Hamiltonians
include non-Hermitian ones, for which the point spectrum sometimes
has a single complex eigenvalue \cite{Fer03,Ros03}. This last
result, combined with appropriate squeezing operators
\cite{Fer00}, could be in connection with the complex-scaling
technique \cite{Ros03}. In general, supersymmetric transformations
constitute a powerful tool in quantum mechanics \cite{Mie04}. As
far as we know, however, the study of Darboux-deformations of
potentials with resonances is still missing in the literature.

In this paper we deal with one-dimensional short range potentials
and their resonances. As usual, the complex energy eigenvalue
$\epsilon = E -i\Gamma/2$ will be taken as a compound of the
resonance $E$ and the inverse of the lifetime $\Gamma$. It will be
shown that resonances can be labelled by a positive integer,
according with their closeness to the scattering threshold. This
integer thus indicates the `grade of excitation' of the related
state: the `ground resonance' will correspond to the least excited
Gamow-Siegert state and so on. We shall use these solutions to
implement Darboux transformations; in this way several exactly
solvable complex and real potentials will be constructed. As we
are going to see, the Darboux-Gamow deformation which is produced
on the initial potential strongly depends on the `excitation' of
the transformation function. Moreover, the complex potentials so
derived behave as an optical device which both refracts and
absorbs light waves. Such a behaviour will be illustrated with the
simplest short range one-dimensional models: square wells and
barriers.

In sections~2 and 3 we present general analytical conditions to
construct Gamow-Siegert functions for one-dimensional short-range
potentials and the corresponding Darboux-deformations. In
Section~4, we will express the transmission coefficient $T$ as a
superposition of Fock-Breit-Wigner (Lorentzian) distributions for
square well potentials. As we shall see, each one of the bell
shaped peaks arises since $T$ has a pole at $\epsilon \in
\mathbb{C}$, which is a complex eigenvalue of the square well. The
Darboux-Gamow deformations of these simple wells give rise to
complex potentials $\widetilde V =\widetilde V_R +i\widetilde V_I$
for which the function $\widetilde V_I$ shows multiple changes of
sign, thus each $\widetilde V$ emits and absorbs flux at the same
time. All the new potentials of this section have
square-integrable functions which represent bound states
associated with eigenvalues in the point spectrum. In Section~5
the case of square barriers is analyzed. We shall show that
Gamow-Siegert functions also lead to a transmission coefficient
which is a sum of Fock-Breit-Wigner distributions. In this case,
`haired' square barriers appear as a consequence of double
Darboux-Gamow deformations for which the number of hairs is in
close connection with the level of excitation of the
transformation function. Finally, in Section~6 we finish the paper
with some concluding remarks. A short appendix is included in
order to present further information which, although important,
can be delayed.


\section{Gamow-Siegert states and one-dimensional short range potentials}

Let us consider the Schr\"odinger equation for a stationary, short
range potential $V(x)$, defined on the straight line $\mathbb{R}$
($\hbar/2m=1$):
\be
Hu(x,\epsilon)=-u''(x,\epsilon)+ V(x) u (x,\epsilon) = \epsilon u
(x,\epsilon).
\label{Schrod1}
\ee
In the sequel, we shall assume that $V(x)$ is characterized by a
strength $V_0$ and a cutoff parameter $\zeta >0$ such that
$V(x)=0$ if $\vert x \vert > \zeta$. A bar over $z\in \mathbb{C}$
stands for complex conjugation ($\overline{z}$) while $z_R$,
$z_I$, denote real and imaginary parts respectively. In general,
$I_+(z)$ denotes the upper half of the $z$-plane ($z_I>0$) and
$I_-(z)$ the lower half plane. Whenever there is no confusion we
use the shortcut notation $f \equiv f(x, \epsilon, \ldots)$,
keeping implicit the dependence of $f$ on $x$, $\epsilon$, and
other possible variables.

It is useful to write the derivative of the logarithm of $u$ as
the complex function $\beta (x, \epsilon)$:
\be
\beta:= -\frac{d}{dx}\ln{u}.
\label{beta}
\ee
Let us also write $u = \Phi e^{i\Xi}$, with $\Phi (x)$ and $\Xi
(x)$ real functions defined as
\be
\Phi = \Phi_0 \exp \left(-\int \beta_R dx\right), \qquad \Xi=
-\int \beta_I dx + \Xi_0
\label{rs}
\ee
where $\Phi_0$ and $\Xi_0$ are integration constants included to
fix the amplitude $\Phi$ and phase $\Xi$ respectively. Thereby,
the current density
\be
j=i\left(\overline{u}'u- \overline{u}u'\right)
\label{corriente}
\ee
can be written as
\be
j= 2 \Xi' \vert u \vert^2 = -2 \beta_I \vert u \vert^2= v \rho.
\label{corrientebeta}
\ee
Hence, $-2 \beta_I$ represents the flux velocity $v$, which plays
an important role in the description of Gamow-Siegert functions as
well as in the Darboux transformations we are going to implement.


\subsection{Asymptotic behaviour of the general solution}

The general solution of (\ref{Schrod1}) for $\vert x \vert >
\zeta$ can be written in terms of ingoing and outgoing waves:
\be
u (x<-\zeta) = I e^{i k x} + L e^{-i k x}, \qquad u (x>\zeta) = N
e^{-i k x} + S e^{ikx}
\label{solgral}
\ee
where coefficients $I,L,N,S,$ depend on the potential parameters
$\zeta$, $V_0$, and the incoming energy $k^2 = \epsilon$. Among
these solutions, we are interested in those which are purely
outgoing waves. The appropriate boundary condition may be written
\be
\lim_{x \rightarrow \pm\infty} \left(u'\mp ik u \right)= \lim_{x
\rightarrow \pm\infty} \{ \left(-\beta \mp ik \right) u \} = 0.
\label{condition}
\ee
Thus, the second term in each of the functions (\ref{solgral})
must dominate over the first one. For such states, equation
(\ref{corriente}) takes the form:
\be
j= \pm(\overline{k}+ k) \vert u \vert^2, \qquad {\rm at} \quad x
\gtrless \pm \zeta.
\label{corriente2}
\ee
If $\epsilon = E$, then $k$ is either pure imaginary or real
according to $E$ is negative or positive. If we assume that the
potential admits negative energies, we get $k_{\pm}= \pm i
\sqrt{\vert E \vert}$ and (\ref{corriente2}) vanishes (the flux
velocity $v$ is zero outside the interaction zone). The solutions
$\psi^{(+)}_E$, connected to $k_+$, are not of our immediate
interest because they are bounded (indeed, the adequate values of
$k_+$ produce square-integrable functions in $\mathbb{R}$). On the
other hand, {\it antibound states\/} $\psi^{(-)}_E$ increase
exponentially as $\vert x \vert \rightarrow +\infty$ and represent
resonances under very special conditions (see e.g. \cite{Bla52}).
However, these states will not be considered here since they do
not exist for arbitrary short range potentials. To exhaust the
cases of a real eigenvalue $\epsilon$, let us take now $E>0$. The
outgoing condition (\ref{condition}) drops the interference term
in the density
\be
\rho (x;t) = \vert N \vert^2 + \vert S \vert^2 + 2 \vert
\overline{N} S \vert \cos (2\kappa x + {\rm Arg} \, S/N), \qquad
x>\zeta,
\label{density}
\ee
so that $\rho = \vert S \vert^2$ is not an integrable function in
neither space nor time (similar expressions hold for $x <-\zeta$).
Remark that flux velocity is not zero outside the interaction zone
($\vert v \vert = 2\sqrt E$). Thereby, $E>0$ provides outgoing
waves at the cost of a net outflow $j \neq 0$. To get solutions
which are more appropriate for this nontrivial $j$, we shall
consider complex eigenvalues $\epsilon$.


\subsection{Resonances}

From the discussion of the previous section it follows that, in
general, $\epsilon$ must be complex. We shall write
\be
\epsilon = E - \frac{i}{2} \, \Gamma, \qquad \epsilon_R =k_R^2
-k_I^2, \qquad \epsilon_I = 2k_R k_I
\label{epsilon}
\ee
where $\epsilon = (k_R +ik_I)^2$. According to (\ref{condition}),
the boundary condition for $\beta$ reads now
\be
\lim_{x \rightarrow \pm \infty} \{ -\beta \pm (k_I -ik_R) \}=0
\label{condbeta}
\ee
so that the flux velocity is $v_+ = + 2 k_R$ for $x > \zeta$ and
$v_- = - 2 k_R$ for $x< -\zeta$. Hence, the ``correct'' direction
in which the outgoing waves move is given by $k_R>0$. In this
case, the density
\be
\rho(x;t) \equiv \vert u(x,t)\vert^2 =  e^{-\Gamma t} \vert
u(x)\vert^2, \qquad \lim_{x\rightarrow \pm \infty} \rho(x;t) =
e^{-\Gamma (t -x/v_\pm)}
\label{damped}
\ee
can be damped by taking $\Gamma>0$. Thereby, $k_I \neq 0$ and $k_R
\neq 0$ have opposite signs. Since $k_R>0$ has been previously
fixed, we have $k_I <0$. Then, purely outgoing, exponentially
increasing functions (resonant states) are defined by points in
the fourth quadrant of the complex $k$-plane. As usual,
$\epsilon_R =E$ will be called a resonance and $\Gamma = 2 \vert
\epsilon_I \vert$ the inverse of lifetime. We shall write $u(x,
\epsilon = E-\frac{i}{2} \Gamma) \equiv \varphi_{\epsilon}(x)$.

Observe that density (\ref{damped}) increases exponentially for
either large $\vert x \vert$ or large negative values of $t$. The
usual interpretation is that the compound ($\varphi_{\epsilon},V$)
represents a decaying system which emitted waves in the remote
past $t-x/v$. As it is well known, the long lifetime limit
($\Gamma \rightarrow 0$) is useful to avoid some of the
complications connected with the limit $t \rightarrow -\infty$
(see discussions on time asymmetry in \cite{Arn99}). In this
context, we shall impose the condition:
\be
\frac{\Gamma/2}{\Delta E}<<1.
\label{vecinas}
\ee
Thus, the level width $\Gamma$ is much smaller than the level
spacing $\Delta E$ in such a way that closer resonances imply
narrower widths (longer lifetimes). An additional condition which
will be useful in the next calculations reads
\be
E> \Delta E.
\label{branch}
\ee

In general, the main difficulty is precisely to find the adequate
$E$ and $\Gamma$. However, for one-dimensional stationary short
range potentials and appropriate values of $V_0$ and $\zeta$, the
transmission coefficient $T$ presents a series of peaks. Each
sharped peak is a bell-shaped curve of width $\Gamma$ which can be
described by the Fock-Breit-Wigner (FBW) distribution
\be
\omega (\epsilon_R, E) = \frac{(\Gamma/2)^2}{(\epsilon_R -E)^2
+(\Gamma/2)^2}.
\label{bw}
\ee
The center $E$ of the peak defines a wave for which the scattering
is ineffective ($E$ is a transparency!). Energies $\epsilon_R$
which are very close to $E$, namely $\vert \epsilon_R -E \vert
\leq \Gamma/2$, induce a resonance: the wave interacts with the
potential in such a way that the time to cross the interaction
zone is maximum. The superposition of a denumerable set of FBW
distributions (each one centered at each resonance $E_n,
n=1,2,\ldots$) entails an approximation of the coefficient $T$
such that the larger the number $N$ of close resonances involved,
the higher the precision of the approximation:
\be
T \approx \omega_N (\epsilon_R) = \sum_{n=1}^N \omega(\epsilon_R,
E_n ).
\label{taprox}
\ee
The above description suggests a graphical method to calculate $E$
and $\Gamma$ (see Appendix~A).


\section{Darboux-Gamow transformations}

In order to throw further light on the complex function $\beta$ we
may note that (\ref{beta}) transforms the Schr\"odinger equation
(\ref{Schrod1}) into a Riccati one
\be
-\beta'+\beta^2+\epsilon=V.
\label{riccati}
\ee
Remark that (\ref{riccati}) is not invariant under a change in the
sign of the superpotential $\beta$
\be
\beta' + \beta^2 + \epsilon= V + 2 \beta'.
\label{nricatti}
\ee
These last equations define a Darboux transformation
$\widetilde{V} \equiv \widetilde{V}(x, \epsilon)$ of the initial
potential $V$. The Darboux-Gamow deformation is necessarily a
complex function
\be
\widetilde{V}=V+2\beta' \equiv V - 2\frac{d^2}{dx^2} \ln
\varphi_{\epsilon}.
\label{Vtilde}
\ee
The solutions $y \equiv y(x,\epsilon,{\cal E})$ of the
non-Hermitian Schr\"odinger equation
\be
-y''+\widetilde{V}y={\cal E} y
\label{Schrod2}
\ee
are easily obtained
\be
y \propto \frac{{\rm
W}(\varphi_{\epsilon},\psi)}{\varphi_{\epsilon}},
\label{1ssol}
\ee
where ${\rm W}(*,*)$ stands for the Wronskian of the involved
functions and $\psi$ is eigensolution of (\ref{Schrod1}) with
eigenvalue ${\cal E}$.

Let us analyze the properties of these new solutions. First, if
$\psi_s$ is the scattering state of $H$ connected with the
positive kinetic parameter $\kappa = \sqrt{{\cal E}_s}$, equations
(\ref{solgral}) become
\be
\psi_{s<} = e^{i\kappa x} + L_s e^{-i\kappa x}, \qquad \psi_{s>} =
S_s e^{i\kappa x}
\label{scat}
\ee
where we have assumed a single source at the left of $V$. Hence,
$T = \vert S_s \vert^2$ and $R = \vert L_s \vert^2$ are such that
$R+T=1$. The transformed scattering waves are obtained from
(\ref{1ssol}) to read
\be
y_{s<} = (\beta_< + i \kappa) e^{i\kappa x} + (\beta_< - i \kappa)
L_s e^{-i\kappa x}, \qquad y_{s>} = (\beta_> + i \kappa) S_s
e^{i\kappa x}.
\label{yscat}
\ee
Hence, we have
\be
\widetilde R = \vert t \vert^2 R, \qquad \widetilde T = \vert t
\vert^2 T, \qquad t= \frac{\beta_> + i\kappa}{\beta_< +i\kappa}.
\label{erete}
\ee
After using (\ref{condbeta}) and considering $\epsilon_R >0$ we
arrive at
\be
\widetilde R + \widetilde T \approx \frac{1+\left( \frac{v_s}{v}
\right)^2 \left[ 1 -2\frac{v}{v_s} \right]}{1+\left( \frac{v_s}{v}
\right)^2 \left[ 1 +2\frac{v}{v_s} \right]}
\label{unotil}
\ee
where $v_s \neq 0$ and $v \neq 0$ are the flux velocities of the
scattering wave $\psi_s$ and the Gamow-Siegert function
$\varphi_{\epsilon}$ respectively. Thereby, $\widetilde R +
\widetilde T \approx 1$, no matter the value of $v_s/v \neq 1$.
That is, scattering waves $\psi_s$ and their Darboux-Gamow
deformations $y_s$ share similar transmission probabilities.

Now, let us suppose that Hamiltonian $H$ includes a point spectrum
$\sigma_d(H) \subset {\rm Sp} (H)$. If $\psi_n$ is a
(square-integrable) eigenfunction with negative eigenvalue ${\cal
E}_n$, then its Darboux-Gamow deformation (\ref{1ssol}) is
bounded:
\be
\lim_{x \rightarrow \pm \infty} y_n = \mp (\sqrt{{\cal E}_n} +
ik)(\lim_{x \rightarrow \pm \infty} \psi_n ).
\label{bound}
\ee
Thereby, $y_n$ is a normalizable eigenfunction of $\widetilde H$
with eigenvalue ${\cal E}_n$. However, as $\epsilon$ is complex,
although the new functions $\{ y_n \}$ may be normalizable, they
will not form an orthogonal set \cite{Ros03} (see also
\cite{Ram03} and the `puzzles' with self orthogonal states
\cite{Sok06}). There is still another bounded solution to be
considered. Function $y_{\epsilon} \propto
\varphi^{-1}_{\epsilon}$ fulfills equation (\ref{Schrod2}) for the
complex eigenvalue $\epsilon$. Since $\lim_{x \rightarrow \, \pm
\infty} \vert y_{\epsilon}\vert^2 = e^{\pm 2 k_I x}$ and $k_I<0$,
we have another normalizable function to be added to the set
$\{y_n\}$.

In summary, one is able to construct non-Hermitian Hamiltonians
$\widetilde{H}$ for which the point spectrum is also
$\sigma_d(H)$, extended by a single complex eigenvalue ${\rm Sp}
(\widetilde{H}) = {\rm Sp} (H) \cup \{\epsilon\}$.


\subsection{New short range real potentials}

The iteration of the previous procedure is easy once a set of
Gamow-Siegert functions $\{ \varphi_{\epsilon} \}$ belonging to
the complex eigenvalues $\{ \epsilon \}$ has been given:
\be
V_2=\widetilde{V}+2\beta_2'(\alpha)
\label{V2}
\ee
where $\beta_2(\alpha)\equiv\beta(x,\epsilon,\alpha)$ is a
B\"acklund-like function \cite{Mie00} defined by the
finite-difference algorithm
\be
\beta_2(\alpha) = -\beta(\epsilon) +  \frac{\alpha
-\epsilon}{\beta(\alpha) - \beta(\epsilon)}
\label{sususy}
\ee
and fulfilling
\be
-\beta_2'(\alpha)+\beta_2^2(\alpha)+\alpha=\widetilde{V},\quad
\alpha\in \mathbb{C}.
\label{2sr}
\ee
In general $V_2 \equiv V_2(x,\epsilon,\alpha)$ is complex;
however, this becomes real if $\alpha=\overline{\epsilon}$
\cite{Fer03,Ros03}:
\be
V_2= V + 2 \left(\frac{\epsilon_I}{\beta_I}\right)' = V - 4
\left(\frac{\epsilon_I}{v}\right)'
\label{V2Real}
\ee
and the new functions $\Psi \equiv \Psi(x, \epsilon, {\cal E})$,
such that $H_2\Psi={\cal E}\Psi$, read
\be
\Psi =(\epsilon - {\cal E}) \psi -2 \left( \frac{\epsilon_I}{v}
\right) y.
\label{2ssol}
\ee
In this case, $V_2$ is a Hermitian short range potential and there
is no additional eigenvalues. Hence, $H_2$ and $H$ are strictly
isospectral. A straightforward calculation shows that
Darboux-Gamow deformations of $\psi_n \in L^2({\mathbb R})$
correspond to normalizable $\Psi_n$ while scattering states
$\psi_s$ are transformed into $\Psi_s$, with similar transmission
probabilities.

In the next sections we shall deform the simplest short range
one-dimensional potentials: square wells and barriers.


\section{Square wells}

Let us consider the square well potential
\be
V(x) = -V_0 \, \Theta \left(\frac{b}{2} - \vert x \vert \right),
\qquad \Theta(x) = \left\{
\begin{array}{l}
1,\quad  x \geq 0\\
0, \quad x<0
\end{array}
\right.
\label{square}
\ee
with $V_0 >0$ and $b>0$. We shall assume that the only source of
particles is at the left of the interaction zone (this
arbitrariness is allowed by the parity symmetry of the potential).
In this way, we shall analyze the scattering by an attractive
potential of range $b$.

The scattering solution of the Schr\"odinger equation
(\ref{Schrod1}) for potential (\ref{square}) reads:
\be
u =\left\{
\begin{array}{cc}
e^{ikx}+ L(k) e^{-ikx},& x< - \frac{b}{2}\\[1.2ex]
\frac{k }{\Delta} \, e^{-ikb/2}  \left[ i(k\cos {\frac{qb}{2}} -iq
\sin{\frac{qb}{2}})
\sin{qx} \right.\\[1ex]
\left. + (q\cos{ \frac{qb}{2}} -ik\sin{\frac{qb}{2}})
\cos{qx} \right], & - \frac{b}{2} \leq  x \leq \frac{b}{2}\\[2ex]
S(k) e^{ikx},& \frac{b}{2} < x
\label{sgral}
\end{array}
\right.
\ee
where
\be
L(k) = i \left( \frac{V_0 \sin qb }{2\Delta} \right) e^{-ikb},
\qquad S(k) = \left( \frac{kq}{\Delta} \right) e^{-ikb}.
\label{s}
\ee
The kinetic and interaction parameters are defined respectively by
$k^2 = \epsilon$ and $q^2 = k^2 + V_0$, while $\Delta$ is the
function
\be
\begin{array}{l}
\Delta(k)= \left(k\cos {\frac{qb}{2}} -iq \sin {\frac{qb}{2}}
\right) \left(q\cos {\frac{qb}{2}} -ik\sin {\frac{qb}{2}}\right).
\label{delta}
\end{array}
\ee
Let us concentrate on the {\it transmission amplitude\/} $S$,
which is regular in $I_+$ except for the points $k_+$ located on
the positive imaginary axis (that is, each $k_+$ corresponds to a
bound state, see Appendix~A). It is simple to verify that $k$ and
$-\overline{k}$ will produce the same peaks in the transmission
coefficient, that is, $\overline{S(k)} =S(-\overline{k})$.
However, they are in $I_-$ (as discussed in Section~2, $k$ is in
the fourth quadrant of the $k$-plane). Hence, $S$ must be extended
to be analytic in $I_-$. First, let us observe that the
relationship
\be
S(-k)S(k)=1
\label{ss}
\ee
is fulfilled for the points $k$ which are such that $qb=n\pi$.
Since $-k \in I_+$, this last equation defines $S(k)=1/S(-k)$ as
an analytical function in $I_-$ but at the points $-k$ which are
zeros of $S$ in $I_+$. In a similar manner, we realize that
$S(-\overline{k})$ is analytical in $I_-$ except for the points
$\bar k$ which are zeros of $S$ in $I_+$. In summary, let $k_n$ be
a pole of $S$ in the fourth quadrant of the $k$-plane, then
$-\overline{k}_n$ is also a pole while $\overline{k}_n$ and $-k_n$
are zeros of $S$. Thereby, $S$ is a meromorphic function of $k$,
with poles restricted to the positive imaginary axis (bound
states) and the lower half-plane (resonances). On the other hand,
if $S$ is studied as a function of $\epsilon$, it is necessary to
consider a two-sheet Riemann surface ($k^2=\epsilon$) with a cut
along the positive real axis. Complex poles of $S(\epsilon)$
always arise in conjugate pairs (corresponding to $k$ and
$-\overline{k}$) while poles on the negative real axis correspond
to either bound or antibound states.


\subsection{Analytical approach for calculating resonances}

In this section we shall derive expressions for the resonance $E$
and the width $\Gamma$ in terms of the potential parameters. We
assume that the incoming kinetic parameter $k$ is a complex number
with nontrivial imaginary part. As the presence of the short range
potential is necessarily encoded in the (complex) interaction
parameter, most of the approximations will be made on $q$.

For complex $q$, additional to (\ref{epsilon}), the real and
imaginary parts of $\epsilon$ read
\be
E = q_R^2 -V_0 - q_I^2, \qquad \frac{\Gamma}{2} = -2q_R q_I.
\label{enq}
\ee
The combination of equations (\ref{enq}) leads to:
\be
\frac{1}{4} \left( \frac{\Gamma}{2} \right)^2 \gamma^2 -(V_0 + E)
\gamma -1 =0
\label{eqgam}
\ee
where $\gamma = q_I^{-2}$. Dropping the negative root we get
\be
q_I^2 = \frac{\frac12 (\Gamma/2)^2}{(V_0 +E)\left( 1 + \sqrt{1 +
\frac{1}{4} \left( \frac{\Gamma/2}{V_0 + E} \right)^2} \right)}.
\label{solq}
\ee
On the other hand, since $V_0 >0$, condition (\ref{branch})
produces
\be
\frac{\Gamma/2}{V_0 + E} < \frac{\Gamma/2}{E} <
\frac{\Gamma/2}{\Delta E}
\label{cadena}
\ee
which, together with condition (\ref{vecinas}), implies
\be
\left( \frac{\Gamma/2}{V_0 + E} \right)^2 \approx 0.
\label{chirris}
\ee
Hence, equation (\ref{solq}) gives
\be
q_I \approx \frac{\Gamma}{4 \sqrt{V_0 +E}}.
\label{chirris2}
\ee
This value of $q_I$, when substituted into the expression for
$\Gamma$ in (\ref{enq}), leads to
\be
E \approx {q_R}^2 - V_0.
\label{mala}
\ee
We first notice that the positiveness of $E$ is ensured whenever
$q_R$ fulfills $q_R^2 > V_0$. Now, a simple comparison between
(\ref{mala}) and (\ref{enq}) shows that $q_I^2$ has to be smaller
than $V_0$. We then have a relation of order $q_I^2 < V_0 <
q_R^2$. As we shall see, it is most convenient to constrain the
values of $q_R$ and $q_I$ as weighted by the cutoff $b/2$:
\be
\left( \frac{q_I b}{2 \theta} \right)^2 < 1 < \left( \frac{q_R
b}{2\theta} \right) ^2, \qquad \theta (V_0, b):= \frac{b}{2}
\sqrt{V_0}
\label{epesado}
\ee
where we have introduced the parameter $\theta$ which encodes not
only the range of application of our approach but the physical
identity of the potential (\ref{square}) as well.

Now, making the r.h.s. factor of (\ref{delta}) equal to zero, we
obtain $q=\pm\sqrt{V_0} \sin \frac{qb}{2}$. After a
straightforward calculation the corresponding equations for the
real and imaginary parts are obtained. For $\theta >>1$ and small
values of $\frac{q_I b}{2}$, the last one leads to $\cos {
\frac{q_R b}{2}} \approx\mp \frac{1}{\theta}$. Therefore, we
arrive at
\be
\frac{q_R b}{2} \approx \frac{n\pi}{2} \pm \frac{1}{\theta} + {\rm
O \left( \frac{1}{\theta^3} \right)}
\label{cota1}
\ee
where $n$ is to be determined. After introducing (\ref{cota1})
into (\ref{mala}), we finally get a discrete set of resonances
\be
E \approx \left[ \left( \frac{n \pi}{2 \theta} \right)^2 -
1\right] V_0, \qquad \frac{n\pi}{2} > \theta, \qquad n\in
\mathbb{N}.
\label{c3}
\ee
It is clear that $n$ must exceed a minimum value to get a positive
energy. Let us take $n := n_{\rm inf} + m$, $m=0,1,\ldots$, where
$n_{\rm inf}$ is the ceiling function of $2\theta/\pi$, i.e.,
$n_{\rm inf} = \left\lceil \frac{2 \theta}{\pi} \right\rceil$. In
this way, $E_{n_{\rm inf}} \equiv E_0$ is the smallest positive
energy in (\ref{c3}). Hence
\be
E_m = \left( \left[ \frac{(n_{\rm inf} + m )\pi}{2 \theta}
\right]^2 - 1\right) V_0, \qquad m \in \mathbb{Z}^+.
\label{c32}
\ee
The derivation of $\Gamma$ is now easy. Making the l.h.s. of
$\Delta$ equal to zero we arrive at $k_I = \pm \sqrt{V_0} \cos
\frac{q_R b}{2} \cosh \frac{q_I b}{2}$. After using (\ref{cota1}),
for small $q_I b/2$ one obtains
\be
-\frac{k_I b}{2} \approx 1 + \frac12 \left(\frac{q_I b}{2}
\right)^2.
\label{kai}
\ee
With this last result and (\ref{mala}) into (\ref{epsilon}), the
energy $E$ can be rewritten as
\be
\left( \frac{k_R b}{2\theta} \right)^2 = \left( \frac{q_R
b}{2\theta} \right)^2 + \left( \frac{q_I b}{2\theta} \right)^2 +
\frac{1}{\theta} -1
\label{choncho}
\ee
the second additive term is smaller than the first one (see
equation (\ref{epesado})) and $\theta^{-1} <<1$, so equation
(\ref{choncho}) reduces to $k_R \approx \pm \sqrt E$. The combined
substitution of these last results into the expression for
$\Gamma$ in (\ref{epsilon}) gives
\be
\frac{\Gamma}{2} \approx \frac{4}{b} \sqrt{ E} = \frac{2v_+}{b}
\label{c5}
\ee
where the terms which are proportional or greater than the square
of $q_I b/2$ have been dropped (compare our results with those
reported in e.g. \cite{Web82}).

It is remarkable that $\Gamma$ is proportional to $v_+$. For a
given value of $b$ ($\theta$ fixed), a hasty incident particle
spreads its peak in $T$ by spending less time in the interaction
zone. As a result, the peaks centered at hot resonant energies
$v^2_+/2$ tend to lose the FBW shape by excessively overlapping
the next ones. The long lifetime limit ($ \Gamma \rightarrow 0$),
applied for fixed values of $\theta$, is then a good criterion to
select velocities $v_+$ as well as potential parameters in order
to obtain appropriate Gamow-Siegert functions for the square
wells.


\subsubsection{Fock-Breit-Wigner distributions}

The above discussion considered a complex kinetic parameter with
nontrivial imaginary part. As we have seen, approximations leading
to $E$ are also useful to get $\Gamma$ in the long lifetime limit.
Next, we are going to verify that $E$ and $\Gamma$ so derived
correspond to the center and width of a FBW peak in the
transmission coefficient $T$. It is convenient to rewrite the
transmission amplitude (\ref{s}) as follows
\be
S(k) = \frac{e^{-ikb}}{\cos qb [ 1 -i g(k)]}, \qquad g(k) = \left(
\frac{k^2 + q^2}{2qk}  \right) \tan qb.
\label{ns}
\ee
Let us consider a small displacement $\delta E$ from the resonance
energy $E_m$ along the real axis of the $\epsilon$-plane. Kinetic
and interaction parameters are mapped as $k \rightarrow k + \delta
k$ and $q \rightarrow q + \frac{1}{2q}\delta E$ respectively. Near
the resonance we get $qb \approx 2n\pi + \frac{b}{2q}\delta E$.
Thus $g \approx (2/\Gamma_m) \delta E$, where we have used
(\ref{c5}). Hence
\[
S(k+\delta k) \approx \frac{e^{-i(k +\delta k - n\pi)b}}{1-i
2\delta E/\Gamma_m}
\]
and finally we get
\be
T(E_m + \delta E) \approx \frac{(\Gamma_m/2)^2}{(\Gamma_m/2)^2 +
(\delta E)^2}.
\label{LBW}
\ee
This last expression is consistent with the previously defined FBW
distribution (\ref{bw}) for small displacements along the real
axis of the $\epsilon$-plane, that is $\mathbb{R} \ni \delta E =
(\epsilon_R - E_m)_{\epsilon_R \rightarrow E_m}$. A comparison of
(\ref{c32}) and (\ref{c5}) with the results obtained from the
graphical method is reported in Table~\ref{tab-pozo} of
Appendix~A.


\subsubsection{Diverse kinds of transformation function}

\begin{figure}[ht]
\centering \epsfig{file=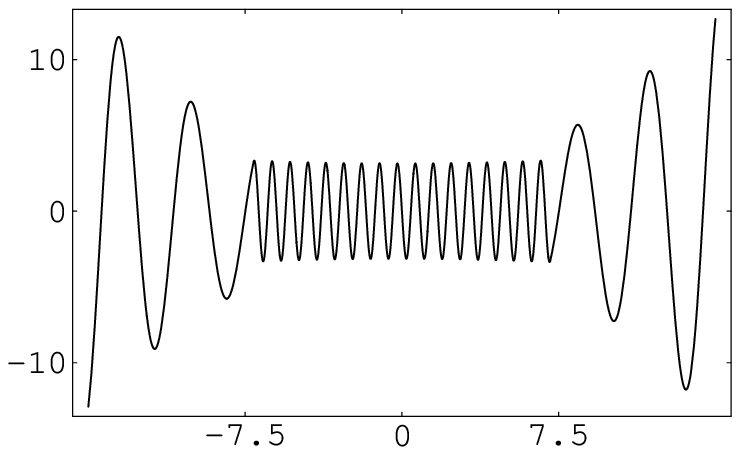, width=5cm} \hspace*{0.5cm}
\epsfig{file=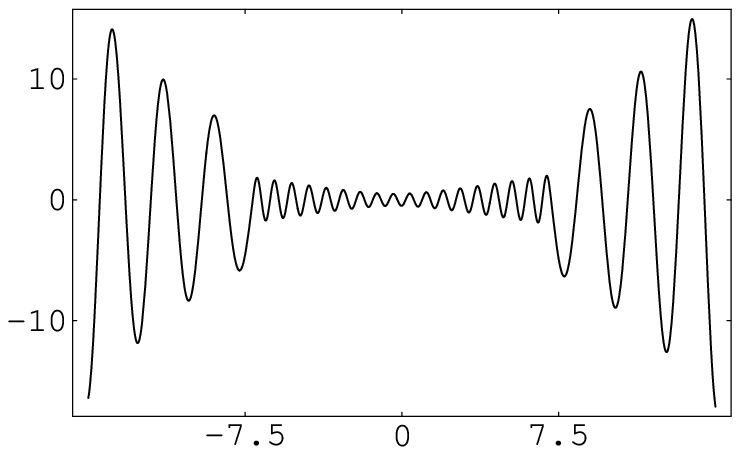, width=5cm}

\caption{\small Real parts of the second (left) and third (right)
decaying states of a square potential defined by $V_0 =50$,
$b=14.2$. The Gamow-Siegert function $\varphi_{\epsilon}$
respectively belongs to the complex eigenvalues $\epsilon_1 =
3.3029 -i 0.5119$ and $\epsilon_2 = 6.5823-i 0.7227$.}
\label{fgorof1}
\end{figure}

\begin{figure}[ht]
\centering \epsfig{file=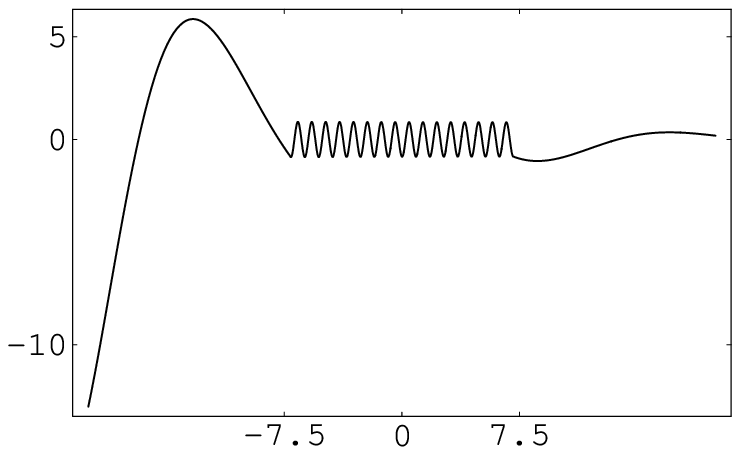, width=5cm} \hspace*{0.5cm}
\epsfig{file=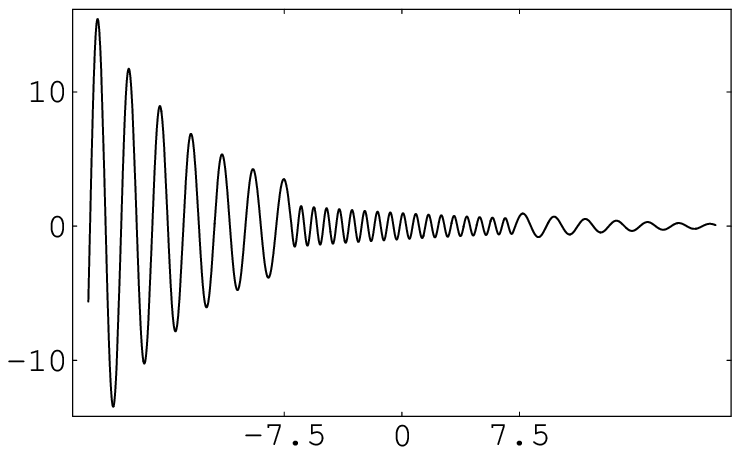, width=5cm}

\caption{\small Imaginary parts of the first (left) and fourth
(right) decreasing functions $\varphi_{\overline{k}}$ of the
potential described in Figure~\ref{fgorof1}. The complex kinetic
parameters are respectively $\overline{k}_0 = 0.3072 + i 0.2484$
and $\overline{k}_3 = 3.1496 + i 0.2811$.}
\label{fgorof2}
\end{figure}

Let us pay attention to solution (\ref{sgral}) for a complex
kinetic parameter $k$. Outside the interaction zone, it reads
\be
u_< = e^{ik_R x} e^{-k_I x} + L(k)\, e^{-ik_R x} e^{k_I x} ,
\qquad u_> = S(k) \, e^{ik_R x} e^{-k_I x}.
\label{gamowb}
\ee
According with the discussion of the above sections, `decaying
states' $\varphi_{\epsilon}(x)$ correspond to kinetic parameters
$k$ living in the fourth quadrant of the complex plane (see
Figure~\ref{fgorof1}). The mirror images in the third quadrant
$-\overline{k}$, give rise to `capture states'
$\varphi_{\overline{\epsilon}}$ behaving in the same global way as
$\varphi_{\epsilon}$ (see equations (\ref{gamowb})). On the other
hand, equations (\ref{gamowb}) evaluated at $\overline{k}$
correspond to an exponentially decreasing function (see
Figure~\ref{fgorof2}). The same is true for $-k$. We shall write
$\varphi_{\overline{k}} \equiv u(x, \overline{k}^2)$. Although
this solution reverts the decaying (\ref{damped}) to an
exponentially growing dependence on time, it is useful for
obtaining special cases of Darboux-Gamow deformations, as we are
going to see.

\subsection{New complex and real wells}

It is common to use complex potentials $V = V_R + i V_I$, $V_I =
{\rm const}$, for describing either absorption or emission of flux
according with the sign of $V_I$ (see e.g. \cite{Fes54}). In
contrast, the case when $V_I$ is an arbitrary function of the
position is rarely analyzed because the related eigenvalue
equation is involved. However, in Section~3 we realized that
Darboux-Gamow deformations $\widetilde V$ are non-trivial exactly
solvable complex potentials.


\begin{figure}[ht]
\centering \epsfig{file=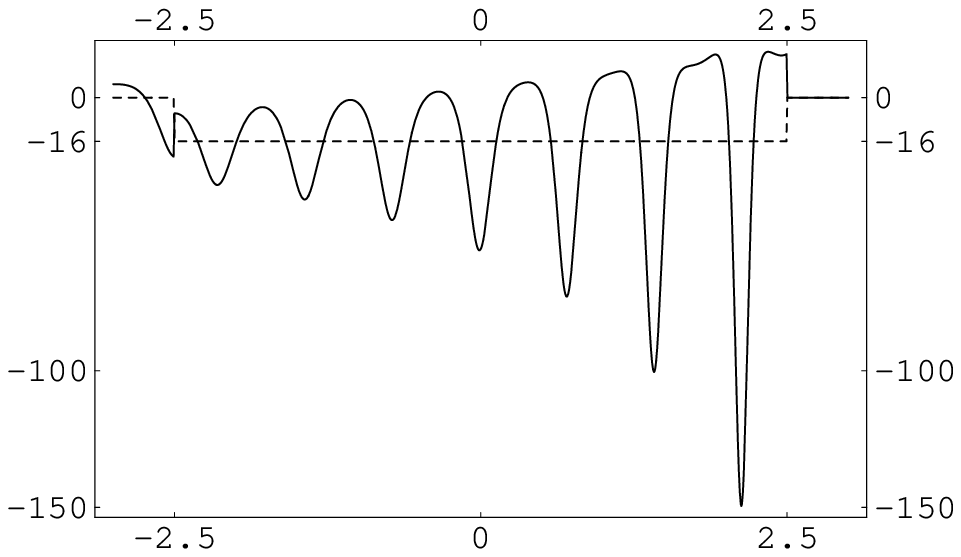, width=5cm} \hspace*{0.5cm}
\epsfig{file=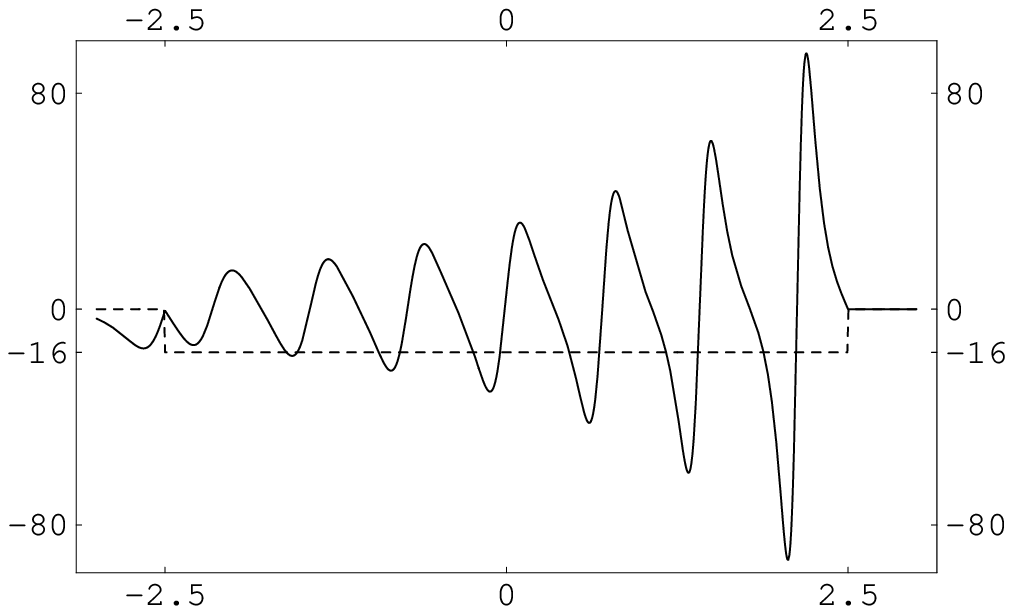, width=5cm}

\caption{\small The real (left) and imaginary (right) parts of the
Darboux-Gamow deformation $\widetilde V$ of a square well (dashed)
for which $V_0=16$ and $b=5$. The lowest decreasing function
$\varphi_{\overline{k}}$ of the initial potential is the
transformation function with $\overline{k}_0 = 1.7504 + i
0.7657$.} \label{fgorof3}
\end{figure}

\begin{figure}[ht]
\centering \epsfig{file=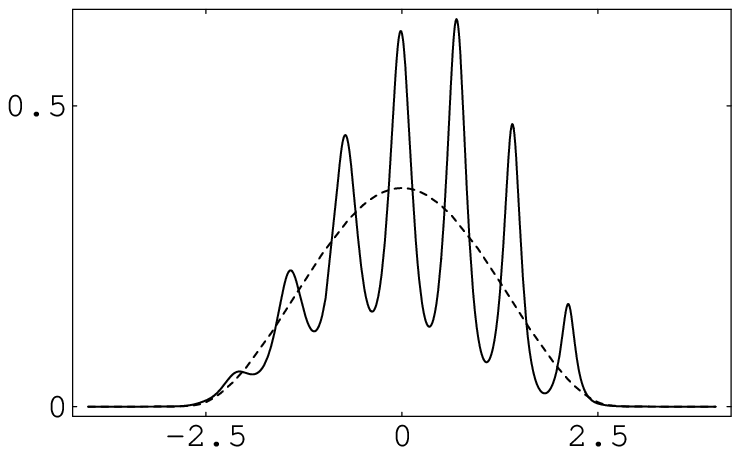, width=5cm} \hspace*{0.5cm}
\epsfig{file=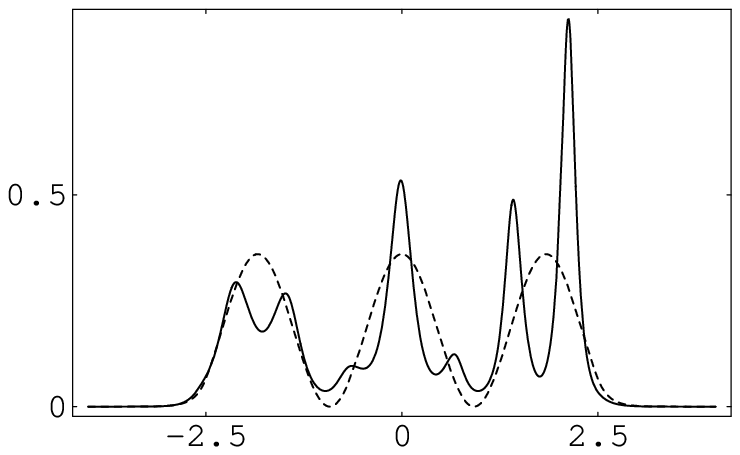, width=5cm}

\caption{\small Left: The probability density belonging to the
normalized ground state (dashed) of a square well potential
($V_0=16$, $b=5$, see Figure~\ref{fap1}) and its Darboux-Gamow
deformation (continuous curve). The transformation parameters are
the same as in Figure~\ref{fgorof3}. Right: The related second
excited states.} \label{fgorof4}
\end{figure}

A typical Darboux-Gamow deformation of the square well
(\ref{square}) is shown in Figure~\ref{fgorof3}. First observe
that the initial parity symmetry is now broken. It is also
remarkable the presence of several local minima and maxima as well
as multiple changes of sign in both functions $\widetilde V_R$ and
$\widetilde V_I$. In a simple optical model the incident beam
crosses a sequence of `obstacles' (potential's maxima and minima),
each one absorbing a fraction of the beam and performing some
operations on the rest in such a way that the states are never
mixed (see discussions on the optical bench in \cite{Mie04}). As a
consequence, the `optical' potential $\widetilde V$ is no longer
self-adjoint and corresponds to an optical device which both
refracts and absorbs light waves. Moreover, this non-Hermitian
potential has a point spectrum and normalized eigenfunctions which
inherit its broken parity symmetry (see Figure~\ref{fgorof4}).


\begin{figure}[ht]
\centering \epsfig{file=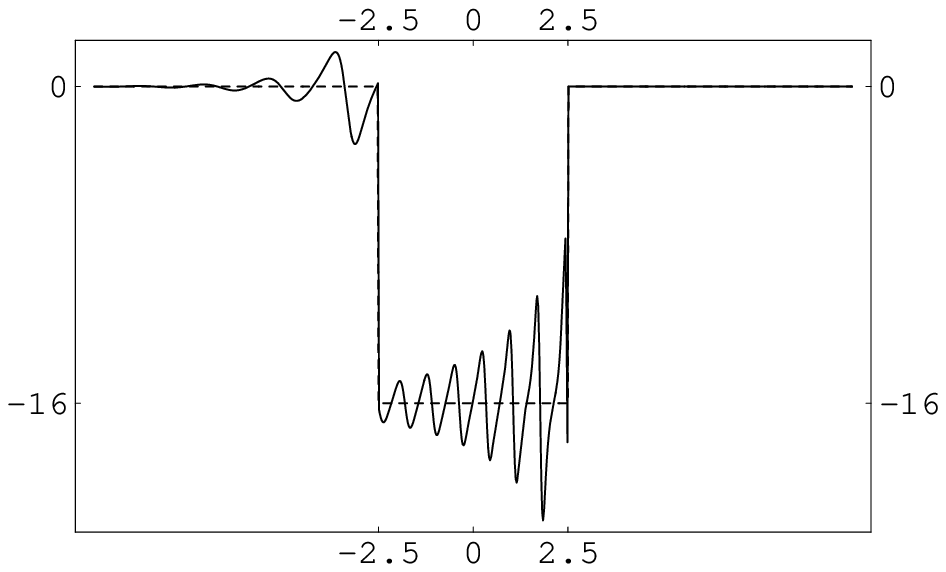, width=5cm} \hspace*{0.5cm}
\epsfig{file=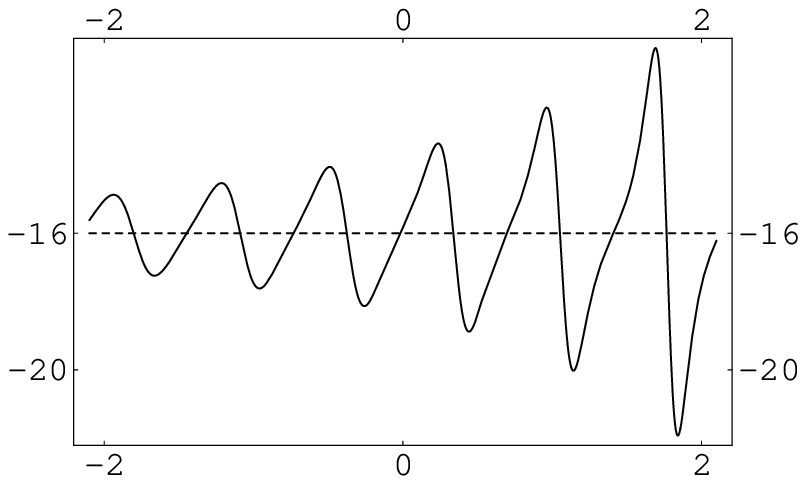, width=5cm}

\caption{\small Twice Darboux-Gamow deformed square well potential
($V_0=16$, $b=5$) for the same transformation parameters as in
Figure~\ref{fgorof3}. A detail of the bottom of this potential is
shown at the right.}
\label{fgorof5}
\end{figure}

On the other hand, new real short range potentials are obtained by
simultaneously using $\varphi_{\epsilon}$ and
$\varphi_{\overline{\epsilon}}$ in (\ref{V2})--(\ref{sususy}). The
result is illustrated in Figure~\ref{fgorof5} with a real
potential $V_2$ for which $V_0 = 16$ and $b=5$. It is notable that
local deformations (that is, the number of maxima and minima)
increases with the level of excitation of the transformation
function.


Finally, let us remark that, although our one-dimensional model is
very simple, extensions to s-waves in three dimensions are
immediate by properly selecting the odd solutions and adding to
the initial potential an impenetrable wall along the negative
axis.


\section{Square barriers}

The barrier potential ($V_0 >0$)
\be
V(x) = V_0 \, \Theta \left( \frac{b}{2} - \vert x \vert \right)
\label{square2}
\ee
is usually studied as a special case of (\ref{square}) for which
the strength $V_0$ is allowed to be a negative number. Thus,
kinetic and interaction parameters now read $k^2 = \epsilon$ and
$q^2 = k^2 -V_0$. The involved solutions are then obtained by
taking $V_0 \rightarrow -V_0$ and $\epsilon = E>0$ in
(\ref{sgral})--(\ref{delta}). However, a simple inspection of
(\ref{c32}) and (\ref{c5}) shows that this is not the case in
analyzing resonances ($E_m$ and $\Gamma_m$ are respectively
negative and pure imaginary numbers if $V_0$ is simply negative!).
In this section we shall obtain the appropriate expressions for
the complex energies (\ref{epsilon}). As in the previous
calculations, we look for the complex roots of $\Delta =0$
fulfilling conditions (\ref{vecinas}) and (\ref{branch}). A
straightforward calculation on the l.h.s. of (\ref{delta}) for the
new $k$ and $q$ leads to
\be
q_R = \pm \sqrt{V_0} \sin \frac{q_R b}{2} \sinh \frac{q_I b}{2},
\qquad q_I = \pm \sqrt{V_0} \cos \frac{q_R b}{2} \cosh \frac{q_I
b}{2}.
\label{qkis}
\ee
The smallness of the weighted $q_I$ is now ensured by taking
\be
\frac{q_R b}{2} = \frac{(2s+1)\pi}{2} + \delta, \qquad
s=0,1,2,\ldots \label{qrre}
\ee
where $\delta$ is a slight perturbation of the weighted $q_R$. The
introduction of (\ref{qkis}) and (\ref{qrre}) into the real and
imaginary parts of $\epsilon$ (compare with equation (\ref{enq}))
gives
\be
E_n \approx V_0 \left[ \left( \frac{(2s+1)\pi}{2}\right)^2 + 1
\right], \qquad \frac{\Gamma_s}{2} = \frac{2}{\theta}(E_s -V_0),
\qquad s=0,1,2,\dots \label{resona}
\ee
In a similar form, the r.h.s. of equation (\ref{delta}) leads to
expressions for $E$ and $\Gamma$ which are the same as
(\ref{resona}) but they hold for semi-integer values of $s$. Thus
we can write $E_n$ and $\Gamma_n$ with $n=1,2,3,\ldots$ such that
$s=(n-1)/2$. Therefore, $E_n>V_0$ for any integer $n \geq 1$.
Moreover, $\Gamma_n$ is always a positive number which grows as $n
\rightarrow \infty$. Thus, resonant behaviour will be better
distinguished, and it will have longer lifetime, for energies
which are close to the top of the barrier. In other words, the
resonance threshold is established by the strength of the
potential (compare our results with those reported in e.g.
\cite{Ant01}). In Table~\ref{tab-barr} of Appendix~A we report a
comparison of the results (\ref{resona}) with those obtained by
means of the graphical method.

\begin{figure}[ht]
\centering \epsfig{file=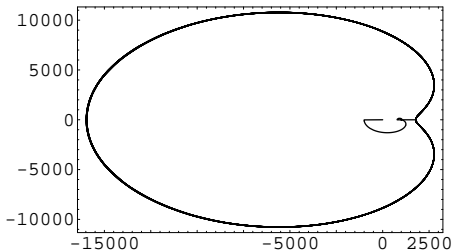, width=6cm}

\caption{\small The Argand-Wessel diagram of the Darboux-Gamow
deformed square barrier for $V_0=1000$, $b=5$ and its first
resonance (see Table~\ref{tab-barr}). The external rings of this
cardiod-like curve correspond to large negative values of the
position $x$.} \label{fgorof6}
\end{figure}

\begin{figure}[ht]
\centering \epsfig{file=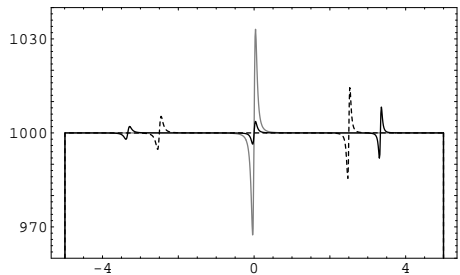, width=6cm}

\caption{\small The top of twice Darboux-Gamow deformations of the
square barrier $V_0=1000$, $b=10$. The lowest decreasing function
produces a very localized distortion (couple of `hairs') on the
new top (gray curve). The second (dotted curve) and third (solid
curve) ones produce respectively two and three distortions (see
Table~\ref{tab-barr} for specific values of the complex
eigenvalues).} \label{fgorof7}
\end{figure}

Figure~\ref{fgorof6} shows the Argand-Wessel diagram of the
Darboux-Gamow deformation of a square barrier. The real and
imaginary parts of this new potential show a series of maxima and
minima which, in a similar manner as for the complex wells, can be
modelled by the optical bench described in \cite{Mie04}. Once the
method is iterated by means of $\varphi_{\overline{\epsilon}}$,
the twice Darboux-Gamow deformed barriers are real. These new
barriers present `hair' over the top which induces stronger
resonant phenomena (see Figure~\ref{fgorof7}). Let us remark that
the number of hairs increases with the excitation of the
transformation function, just as it occurs for the nodes in
bounded wavefunctions. Thus, the lowest resonance induces only one
very localized distortion (a couple of hairs combed in opposite
direction), the second one induces four hairs and so on. Observe
also the global asymmetry of the new barriers, that is, $V_2(-x)
\neq V_2(x)$. Finally, it is reasonable to assume that these
haired barriers will induce delays on the scattering states which
are longer than the delay produced by conventional square barriers
(see simple models in \cite{Mos51}). The analysis of this
phenomenon will be published elsewhere.

\section{Concluding remarks}

We have derived analytical expressions for the resonant energies
of short range one-dimensional potentials in the long lifetime
limit of the resonance levels. The same calculations leading to
the `quantization' $E_m$ of the resonances are appropriate to
obtain the corresponding lifetime inverses $\Gamma_m$. These
results allowed the description of the transmission coefficient as
a superposition of Fock-Breit-Wigner distributions, which is as
good as large is the number of resonances involved. Moreover, they
also were used to transform Hermitian short range potentials into
non-Hermitian ones which either preserve the initial energy
spectra or include a complex eigenvalue with square-integrable
eigenfunction. The new complex potentials are `opaque' in the
sense that they simultaneously emit and absorbs flux, just as an
optical device which both refracts and absorbs light waves. The
iteration of the method ($\epsilon$ and $\overline{\epsilon}$
combined) produces new real potentials for which the energy
spectrum is exactly the same as the initial one.

The Darboux-Gamow deformations of square wells and square barriers
have been presented as straightforward applications. In the firs
case, it has been shown that square-integrable eigenfunctions are
mapped into square-integrable deformed eigenfunctions. In the
second case we obtained `haired' square barriers as a consequence
of double Darboux-Gamow transformations. The hairs are very
localized deformations on the top of the new barriers and are such
that their number depends directly on the excitation of the
involved transformation function. In all cases, the scattering
states preserve their global properties after the deformations.

Some final comments would be important. Processes in which the
incident wave falls upon a single scatterer are fundamental in the
study of more involved interactions \cite{Nus72}. In general, for
a single target the scattering amplitude is a function of two
variables (e.g. energy and angular momentum). Our model
corresponds to the situation in which one of the variables is held
fixed (namely, the angular momentum). A more realistic three
dimensional model is easily obtained from our results: even
functions are dropped while an infinitely extended, impenetrable
wall is added at the negative part of the straight line. Such a
situation corresponds to $s$-waves interacting with a single,
spherically symmetric, square scatterer (see e.g. \cite{Fes54}).
On the other hand, the production of confined states within the
continuum associated with lattice impurity (contact effect) has
been recently reported in the context of Darboux-deformations
\cite{Fer02}. Such a result indicates that our Darboux-Gamow
approach can be extended to the case of spatially oscillating
square well potentials. Then, the presence of `localized' lattice
deformations produced by resonant states could be in connection
with the isolated transitions, observed in semiconductors, from a
bound state within a quantum well to a bound state at an energy
greater than the barrier height \cite{Cap92}. Work in this
direction is in progress.


\section*{Acknowledgements}

The support of CONACyT projects 24233-50766 and 49253-F is
acknowledged.



\section*{Appendix A.}

\begin{table}[ht]
\centering

{\footnotesize


\begin{tabular}{|l|l|c|}
\hline \multicolumn{3}{|c|}{$V_0=1000,\quad b=20,
\quad n_{\rm inf}=202, \quad \epsilon = E_m -i\Gamma_m/2$}\\
\hline
\multicolumn{1}{|c|}{Graphic} & \multicolumn{2}{|c|}{Analytical} \\
\hline
$06.798680-i0.527888$ & $06.798344-i0.521472$ &  $m=0$\\
\hline
$16.790717-i0.843441$ & $16.791319-i0.819544$ & $m=1$\\
\hline
$26.832718-i1.084738$ & $26.833641-i1.036023$ & $m=2$\\
\hline
$36.926378-i1.295156$ & $36.925312-i1.215324$ & $m=3$\\
\hline
$47.065373-i1.489006$ & $47.066330-i1.372098$ & $m=4$\\
\hline
$57.258014-i1.673245$ & $57.256697-i1.513363$ & $m=5$\\
\hline
$67.497189-i1.852008$ & $67.496412-i1.643124$ & $m=6$\\
\hline
$77.786734-i2.028109$ & $77.785474-i1.763421$ & $m=7$\\
\hline
\multicolumn{3}{|c|}{}\\
\hline \multicolumn{3}{|c|}{$V_0=743, \quad b=22, \quad
n_{\rm inf}=191, \quad \epsilon = E_m -i\Gamma_m/2$}\\
\hline
\multicolumn{1}{|c|}{Graphic} & \multicolumn{2}{|c|}{Analytical} \\
\hline
$00.911072-i0.174723$ & $00.911235-i0.173561$ & $m=0$\\
\hline
$08.720839-i0.547966$ & $08.721274-i0.536941$ & $m=1$\\
\hline
$16.571405-i0.768964$ & $16.572095-i0.740160$ & $m=2$\\
\hline
$24.463107-i0.951658$ & $24.463700-i0.899287$ & $m=3$\\
\hline
$32.397141-i1.116079$ & $32.396089-i1.034864$ & $m=4$\\
\hline
$40.370347-i1.270414$ & $40.369261-i1.155214$ & $m=5$\\
\hline
$48.381943-i1.419086$ & $48.383217-i1.264691$ & $m=6$\\
\hline
$56.436735-i1.564930$ & $56.437956-i1.365912$ & $m=7$\\
\hline
\end{tabular}
}

\caption{\footnotesize Representative numerical results for the
resonance energies and widths of a square well. The potential
parameters have been chosen such that the function
$\omega_N(\epsilon_R)$ matches the transmission coefficient $T$,
at least for the first 8 peaks.} \label{tab-pozo}
\end{table}



\begin{table}[ht]
\centering

{\footnotesize

\begin{tabular}{|l|l|c|}
\hline
\multicolumn{3}{|c|}{$V_0=1000, \quad b=5, \quad \epsilon = E_n -i\Gamma_n/2$}\\
\hline
\multicolumn{1}{|c|}{Graphic} & \multicolumn{2}{|c|}{Analytical} \\
\hline
$1000.394784 -i 0.009995$ & $1000.394784-i0.009987$ & $n=1$\\
\hline
$1001.579136-i0.040036$ & $1001.579136-i0.039949$ & $n=2$\\
\hline
$1003.553057-i0.090291$ & $1003.553057-i0.089886$ & $n=3$\\
\hline
$1006.316546-i0.161043$ & $1006.316546-i0.159797$ & $n=4$\\
\hline
$1009.869604-i0.252698$ & $1009.869604-i0.249683$ & $n=5$\\
\hline
$1014.212230-i0.365794$ & $1014.212230-i0.359544$ & $n=6$\\
\hline
$1019.344424-i0.501014$ & $1019.344424-i0.489379$ & $n=7$\\
\hline
$1025.266187-i0.659204$ & $1025.266187-i0.639189$ & $n=8$\\
\hline
\multicolumn{3}{|c|}{}\\
\hline
\multicolumn{3}{|c|}{$V_0=1000, \quad b=10, \quad \epsilon = E_n -i\Gamma_n/2$}\\
\hline
\multicolumn{1}{|c|}{Graphic}  & \multicolumn{2}{|c|}{Analytical} \\
\hline
$1000.098696-i0.001248$ & $1000.098696-i0.001248$ & $n=1$\\
\hline
$1000.394784-i0.004996$ & $1000.394784-i0.004993$ & $n=2$\\
\hline
$1000.888264-i0.011248$ & $1000.888264-i0.011235$ & $n=3$\\
\hline
$1001.579136-i0.020013$ & $1001.579136-i0.019974$ & $n=4$\\
\hline
$1002.467401-i0.031303$ & $1002.467401-i0.031210$ & $n=5$\\
\hline
$1003.553057-i0.045134$ & $1003.553057-i0.044943$ & $n=6$\\
\hline
$1004.836106-i0.061525$ & $1004.836106-i0.061172$ & $n=7$\\
\hline
$1006.316546-i0.080501$ & $1006.316546-i0.079898$ & $n=8$\\
\hline
\end{tabular}

}

\caption{\footnotesize Representative numerical results for the
resonance energies and widths of a square barrier. The potential
parameters have been chosen such that the function
$\omega_N(\epsilon_R)$ matches well the transmission coefficient
$T$, at least for the first 8 peaks.} \label{tab-barr}
\end{table}

\begin{figure}[ht]
\centering \epsfig{file=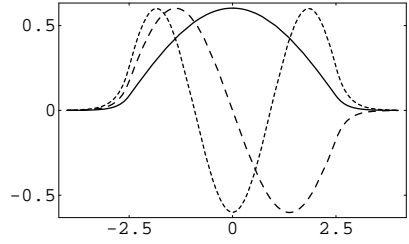, width=6cm}

\caption{\small The ground state (continuous curve) and the first
two excited bound states of a square well $V_0 =16$, $b=5$. The
energies are respectively ${\cal E}_0=-15.6379$, ${\cal E}
_1=-14.6983$, ${\cal E}_2=-13.0812$.}
\label{fap1}
\end{figure}

The physical properties of the solution (\ref{sgral}) are encoded
in $\Delta$. For example, the product $u \Delta$ makes evident
that, after suppressing the incoming wave (i.e., $\Delta =0$),
still there are `reflected' and `transmitted' waves as well as a
finite wave inside the interaction zone. Thus, the outgoing
condition (\ref{condition}) is automatically satisfied. Moreover,
since specific values of $k_+ =i \sqrt{\vert E \vert}$ produce
both the quantization of $E<0$ and the vanishing of the related
solutions at $\pm \infty$, the usual interpretation of bound
states as resonances with infinite lifetime ($\Gamma =0$) is then
recovered. Conventional expressions for these solutions are
usually chosen to be simultaneously eigenstates of the parity
operator, i.e. odd and even functions:
\be
\psi_{\rm odd}(x) = \left\{
\begin{array}{cc}
\exp(\sqrt{\vert E \vert}\, x), & x<-\frac{b}{2}\\[1.5ex]
-\left( \frac{\exp(-\frac{b}{2}\sqrt{\vert E \vert})}{\sin
\frac{qb}{2}} \right) \sin qx, & -\frac{b}{2} \leq x \leq \frac{b}{2}\\[2ex]
-\exp(-\sqrt{\vert E \vert} \, x), & \frac{b}{2}<x
\end{array}
\right. \qquad
\label{odd}
\ee

\be
\psi_{\rm even}(x) = \left\{
\begin{array}{cc}
\exp(\sqrt{\vert E \vert}\, x), & x<-\frac{b}{2}\\[1.5ex]
\left( \frac{\exp(-\frac{b}{2}\sqrt{\vert E \vert})}{\cos
\frac{qb}{2}} \right) \cos qx, & -\frac{b}{2} \leq x \leq \frac{b}{2}\\[2ex]
\exp(-\sqrt{\vert E \vert} \, x), & \frac{b}{2}<x
\end{array}
\right.
\label{even}
\ee
The energy eigenvalue conditions for these solutions are
respectively
\be
\cot \varrho = - \frac{\sqrt{\theta^2 -\varrho^2}}{\varrho},
\qquad \tan \varrho =  \frac{\sqrt{\theta^2 -\varrho^2}}{\varrho}
\label{encond}
\ee
with $\theta$ defined in (\ref{epesado}) and $\varrho = qb/2$.
Figure~\ref{fap1} shows some of the first normalized
eigenfunctions we are dealing with.

\begin{figure}[ht]
\centering \epsfig{file=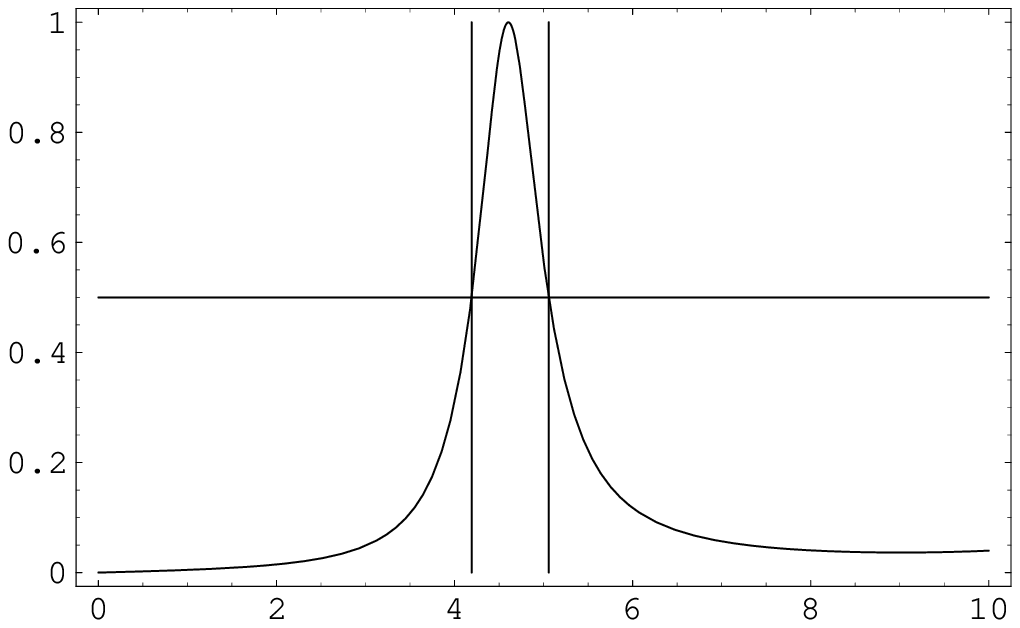, width=5cm} \hskip1cm
\epsfig{file=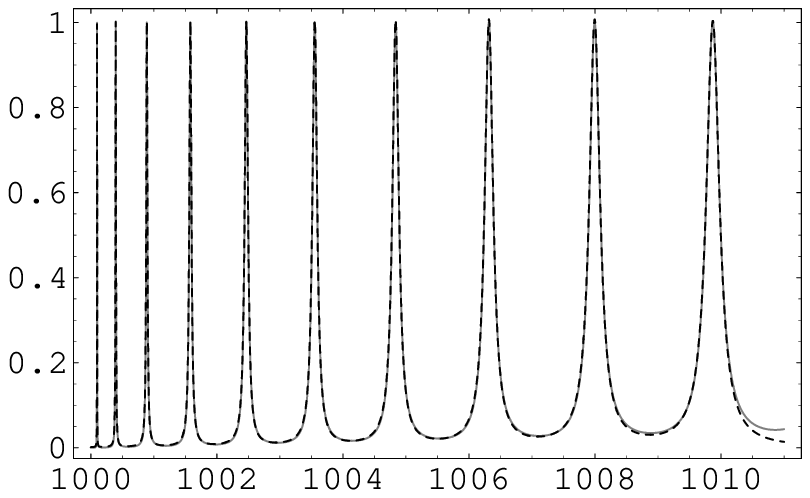, width=5cm}

\caption{\small Left: First resonance of the square well $V_0=
992.25$, $b=20$. The graphic definition of the width $\Gamma$
corresponds to the distance between the vertical lines centering
the peak. Right: Functions $T$ and $\omega_N$ (dotted curve) for
the square barrier $V_0 =10^3$, $b=10$. In this case, the FBW sum
matches well the transmission coefficient for the first ten
resonances (see Table~\ref{tab-barr}).} \label{fa1}
\end{figure}

On the other hand, for one-dimensional short range potentials the
transmission amplitude $S$ could be interpreted as a $1\times1$
scattering matrix. So, to measure the cross section as a function
of the energy of incoming particles corresponds to identify the
local maxima of the transmission coefficient $T$. First, notice
that a sharped peak of $T$ can be connected with a
Fock-Breit-Wigner distribution (\ref{bw}), just as it is
illustrated in Figure~\ref{fa1}. The peaks surrounded by a couple
of local minima which are above $1/2$ will be dropped. The top of
each of the included peaks, projected on the energy axis, defines
a resonance. The width $\Gamma$ of each peak is then defined by
the distance between its right and left intersections with the
horizontal at $1/2$. Function $\omega_N (\epsilon_R)$ in
(\ref{taprox}) includes $N$ of these peaks, counted from left to
right. In Figure~\ref{fa1} it is also shown a case for which the
matching between $T$ and $\omega_N$ is very good for the first ten
resonances. Finally, Table~\ref{tab-pozo} shows the results
obtained with the graphical method as compared with the analytical
procedure of Section~4 for the square well and different values of
the range $b$ and strength $V_0$. Table~\ref{tab-barr} shows the
corresponding results for the square barrier of Section~5.

\vfill



\end{document}